\documentclass[%
 reprint,
 superscriptaddress,
 amsmath,amssymb,
 aps,
 prb,
]{revtex4-2}

\usepackage{graphicx}
\usepackage{dcolumn}
\usepackage{bm}
\usepackage{hyperref}

\begin{document}

\preprint{APS/123-QED}

\title{Maximal Entanglement and Frozen Information: A Unified Framework for Dynamical Quantum Phase Transitions}

\author{Kaiyuan Cao}
 \email{kycao@yzu.edu.cn}
\affiliation{College of Physics Science and Technology, Yangzhou University, Yangzhou 225009, People's Republic of China}

\author{Mingzhi Li}
\affiliation{College of Physics Science and Technology, Yangzhou University, Yangzhou 225009, People's Republic of China}

\author{Xiang-Ping Jiang}
\affiliation{School of Physics, Hangzhou Normal University, Hangzhou, Zhejiang 311121, China}

\author{Shu Chen}
 \email{schen@iphy.ac.cn}
\affiliation{Beijing National Laboratory for Condensed Matter Physics, Institute of Physics, Chinese Academy of Sciences, Beijing 100190, China}

\author{Jian Wang}
 \email{phcwj@hotmail.com}
\affiliation{College of Physics Science and Technology, Yangzhou University, Yangzhou 225009, People's Republic of China}

\date{\today}

\begin{abstract}
    Dynamical quantum phase transitions (DQPTs) are temporal singularities marked by zeros of the Loschmidt echo, yet their underlying quantum-information structure remains elusive. Here, we introduce a momentum-resolved entanglement entropy as a direct probe of DQPTs in translation-invariant free systems. We analytically establish that every critical momentum mode $k^{*}$ associated with a DQPT saturates its entanglement to the maximal value $\ln{2}$, coinciding with the vanishing of the Loschmidt echo. Crucially, we demonstrate that this maximal entanglement universally suppresses information scrambling: a momentum-resolved out-of-time-ordered correlator (OTOC) vanishes identically for all times at $k^{*}$. These three signatures --- Fisher zeros, maximal entanglement, and vanished OTOC --- are proved to be equivalent in both the transverse-field Ising and Su-Schrieffer-Heeger models, despite their distinct bipartitions (momentum-pair vs. sublattice). Our results establish a unified, information-theoretic framework for DQPTs, revealing them a points where quantum correlations saturate and information flow halts. This work elevates entanglement and scrambling to central dynamical order parameters, offering a universal perspective on nonequilibrium quantum critically.
\end{abstract}

\maketitle


\textit{Introduction.---}
The study of quantum phase transitions has been profoundly transformed by the application of entanglement theory. In equilibrium, the entanglement entropy has emerged as a universal diagnostic tool, capable of characterizing diverse phenomena such as topological order, symmetry-breaking criticality, and the underlying conformal field theories \cite{Amico2008rmp, Pasquale2004jsm, Kitaev2006prl, Levin2006prl}. This success establishes a powerful paradigm: entanglement measures serve as fundamental order parameters for equilibrium quantum criticality.

Dynamical quantum phase transitions (DQPTs) extend the notion of criticality to the time domain, marked by nonanalyticities in the Loschmidt echo following a quantum quench \cite{Heyl2013prl, Zvyagin2016ltp, Heyl2018rpp}. So far, DQPTs have been extensively studied in translation-invariant systems both theoretically in clean spin chains \cite{Vajna2014prb, Cao2022cpb, Porta2020scirep, Karrasch2013prb, Zhou2021prb} and fermionic models \cite{Sharma2015prb, Cao2024jpcm, Cao2025pra}, and experimentally in platforms like optical lattices and quantum walks that offer natural access to momentum-space observables \cite{Vogel2017naturep, Chen2020pra, Nie2020prl, Wang2019prl, Xu2020lightsa}. Despite these progresses, a parallel information-theoretically understanding of DQPTs remains underdeveloped. While the Loschmidt echo robustly signals these dynamical singularities, it provides a global, integrated measure that conceals the mode-resolved quantum-informational structure underlying the transition. Consequently, a fundamental gap persists: What universal reorganization of quantum correlations and information flow defines a DQPT, and how can we resolve it in the natural, momentum-space basis of the dynamics? Bridging this gap requires a new framework that elevates information-theoretic quantities from passive observables to central, mode-resolved order parameters for dynamical criticality.

Recent advances have begun to link DQPTs to entanglement in real space \cite{Jurcevic2017prl, Torlai2014jsm, Canovi2014prb, Nicola2021prl, Wong2024prb}. These works have provided valuable insights, observing phenomena such as extrema of enhanced growth of real-space entanglement entropy near critical times \cite{Nicola2021prl}, and the vanishing of Schmidt gaps in the entanglement spectrum at DQPT \cite{Torlai2014jsm, Canovi2014prb}. Furthermore, maximal and minimal entanglement between specific sublattice-pair has been shown to coincide with DQPT in certain settings \cite{Wong2024prb}. However, these findings, while suggestive, remain fragmented and mode-specific. They have not established a universal principle that directly links the defining feature of a DQPT. Crucially, the implications of such critical entanglement for the dynamical flow of quantum information (e.g., scrambling) remain entirely unexplored.

In a recent preprint \cite{cao2026}, we reported a surprising and exact correspondence in the XY chain under squeezing: at the critical momentum of a DQPT, the entanglement entropy between $(k, -k)$ modes saturates to its maximum, $\ln{2}$. This discovery hinted at a deeper universality. In the present work, we substantially generalize and formalize this observation into a complete information-theoretic framework. We demonstrate that the saturation of momentum-space entanglement is not merely a coincidence but is fundamentally equivalent to the complete suppression of information scrambling within the critically entangled sector, as diagnosed by a vanishing momentum-resolved out-of-time-ordered correlator (OTOC). This triad of equivalences --- Fisher zeros, maximal entanglement, and vanished scrambling --- is proved to hold universally across different models and bipartitions, thereby resolving the previously unclear principle underlying the entanglement-DQPT connection.

\textit{Theoretical Framework: A Mode-Resolved Information-Theoretic Characterization of DQPTs---}
The dynamics of translation-invariant free-particle systems following a global quench naturally factorizes into independent momentum sectors. This factorization provides a privileged basis for resolving quantum correlations and information flow in the system's intrinsic degrees of freedom. Here, we formalized the mode-resolved information-theoretic framework proposed in the Introduction. We introduce two complementary quantities that probe, respectively, the entanglement entropy and OTOC within each naturally defined subsystem at momentum $k$.
\begin{enumerate}
    \item \textbf{Momentum-space entanglement entropy}: Entanglement entropy has been established as a fundamental measure of quantum correlations in many-body systems both in equilibrium \cite{Amico2008rmp, Eisert2010rmp} and out of equilibrium \cite{Pasquale2004jsm}. To resolve these correlations in the natural basis of translation-invariant dynamics, we define a momentum-space entanglement entropy \cite{Alves2019epjc, Lundgren2016prb, Dora2016prl, Flynn2023prb}. Consider a translation-invariant system of free fermions or bosons quenched from an initial state $|\Psi(0)\rangle$. The time-evolved pure state factorizes as $|\Psi(t)\rangle = \bigotimes_{k}|\psi_{k}(t)\rangle$, where $k$ runs over independent momentum sectors. For each sector --- which may represent a pair of conjugate model $(k,-k)$ or internal degrees of freedom at the same $k$ --- we define a bipartition according to the natural structure of the dynamics. Tracing out one subsystem yields a reduced density matrix $\rho_{k}(t)$ for the complementary subsystem. The associated momentum-space entanglement entropy is then defined as 
    \begin{equation}
        \mathcal{S}_{k}(t) = -\operatorname{Tr}\bigl[\rho_{k}(t)\ln{\rho_{k}(t)}\bigr].
    \end{equation}
    For the effective two-level system arising in the models studied here, the maximum possible value is $\mathcal{S}_{k, \mathrm{max}} = \ln{2}$ \cite{Vidal2003prl, Nielsen2010book}, corresponding to a maximally mixed reduced state --- a signature of maximal quantum correlation between the two subsystems at momentum $k$.

    \item \textbf{Momentum-space OTOC}: While $\mathcal{S}_{k}(t)$ quantifies the instantaneous quantum correlation, understanding how information spreads and scrambles within correlated sector requires a dynamical measure. The OTOC has emerged as a central diagnostic for quantum chaos and information scrambling \cite{Maldacena2016jhep, Swingle2016pra, Li2020prr}. To resolve this dynamics in momentum space, we introduce momentum-space OTOC, tailored to the same bipartition used for $\mathcal{S}_{k}(t)$. Let $V_{k}$ and $W_{k}$ be Hermitian operators acting on the two subsystem of the chosen bipartition. The OTOC is defined as \cite{Maldacena2016jhep}
    \begin{equation}\label{eq: otoc.definition}
        C_{k}(t) = -\langle\Psi(0)|[W_{k}(t), V_{k}]^{2}|\Psi(0)\rangle,
    \end{equation}
    where $W_{k}(t)$ is the Heisenberg-evolved operator. For concreteness, we will choose $V_{k}$ and $W_{k}$ to be local occupation-number operators (e.g., $n_{k}$ and $n_{-k}$ for a pairing model, or $n_{Ak}$ and $n_{Bk}$ for a two-sublattice model). This choice directly probes how a local perturbation in one subsystem propagates to affect the other within the same momentum sector, extending the concept of operator spreading to the mode-resolved context \cite{Swingle2016pra}.

    \item \textbf{The unified conjecture --- equivalence of critical signatures}: The Loschmidt amplitude, which robustly signals DQPTs via its nonanalyticities, factorizes into independent momentum sectors in translation-invariant systems: $\mathcal{G}(t) = \prod_{k}\mathcal{G}_{k}(t)$ \cite{Heyl2013prl, Budich2016prb}. Crucially, the overall rate function becomes nonanalytic if and only if at least one critical mode $k^{*}$ contributes a real Fisher zero, $\mathcal{G}_{k^{*}}(t) = 0$. This establishes the momentum mode as the fundamental unit of dynamical criticality. However, while the global Loschmidt echo pinpoints the occurrence of a DQPT, it provides an integrated, singular measure that conceals the intrinsic quantum information structure of the critical mode itself.

    We now formulate the central conjecture of our information-theoretic framework to resolve this structure. In translation-invariant free-particle systems, the following three conditions at critical momentum $k^{*}$ are equivalent, arising from the same underlying critical condition on quench parameters:
    \begin{enumerate}
        \item Fisher zeros: $\mathcal{G}_{k^{*}}(t^{*}) = 0$,
        \item Maximal mode entanglement: $\mathcal{S}_{k^{*}}(t) = \mathcal{S}_{k,\mathrm{max}} = \ln{2}$,
        \item Vanished mode scrambling: $C_{k^{*}}(t) \equiv 0$ for all times $t$.
    \end{enumerate}
\end{enumerate}
The profound implication of the equivalence between (a) and (b) is a complete restructuring of information dynamics at the critical mode. Maximum entanglement implies the reduced state is maximally mixed ($\rho = I/d$). In such a state, any local perturbation is ``absorbed'' into the pre-existing maximal randomness, preventing its detectable propagation to the partner subsystem. Consequently, the Heisenberg-evolved operator $W_{k^{*}}(t)$ commutes with $V_{k^{*}}$, leading to the identical vanishing of the OTOC \cite{Li2020prr}. Thus, a DQPT marks a point where, for the critical mode, quantum correlations saturate and information propagation freezes --- the system reaches a dynamical fixed point in information space.

\begin{figure}
    \centering
    \includegraphics[width=1\linewidth]{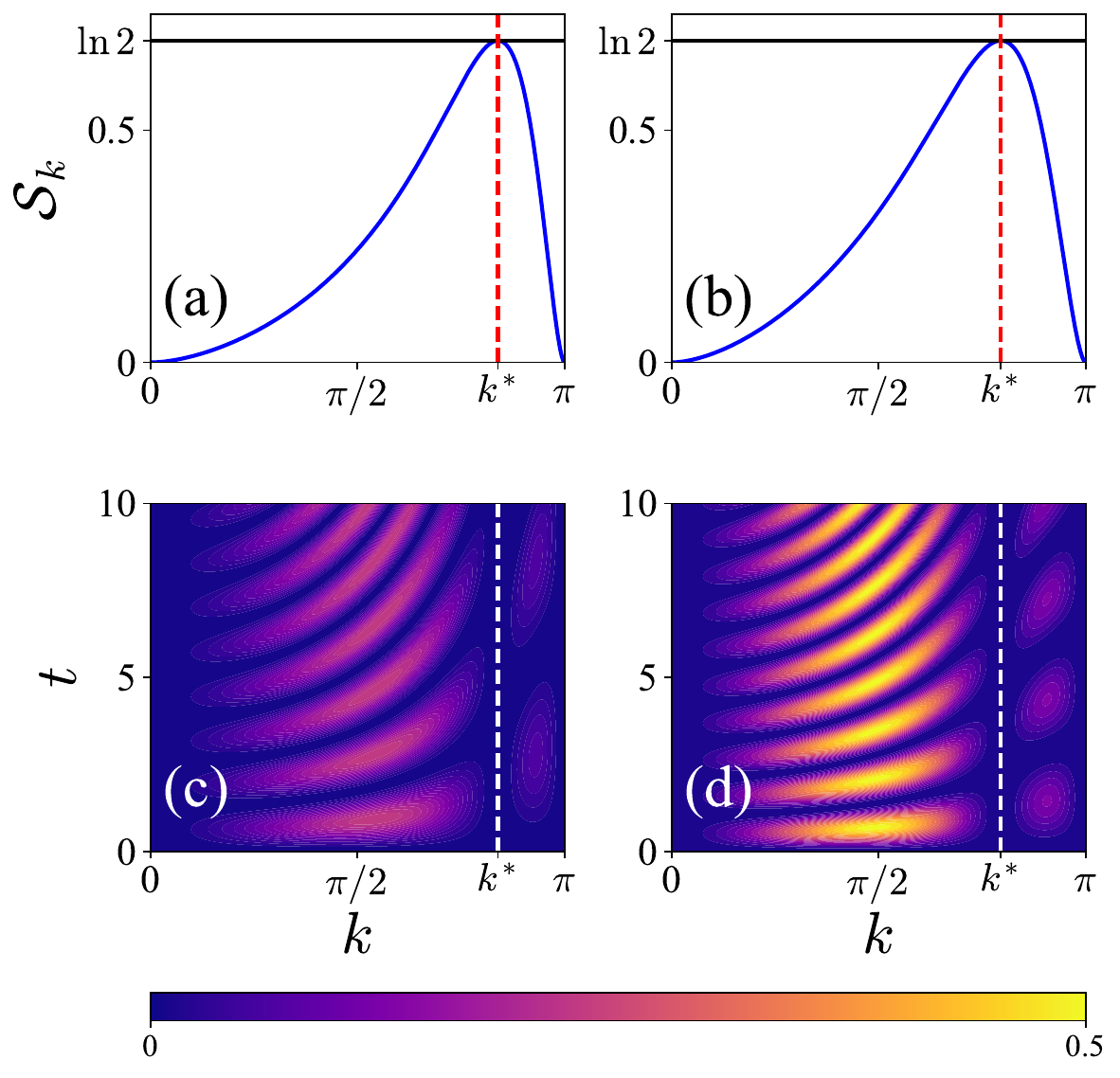}
    \caption{(a) (c) The momentum-space entanglement entropy $\mathcal{S}_{k}$ and OTOC $C_{k}(t) = -\langle [n_{k}(t), n_{-k}]^{2} \rangle$ in the TFI chain for a quench from $h_{0}=0.5$ to $h_{1}=1.5$. (b) (d) Sublattice momentum-space entanglement entropy $\mathcal{S}_{k}$ and OTOC $C_{k}(t) = -\langle [n_{Ak}(t), n_{Bk}]^{2} \rangle$ in the SSH model for a quench from $h_{0}=0.5$ to $h_{1}=2$.}
    \label{fig: entropy.otoc}
\end{figure}

In the following sections, we prove this conjecture explicitly in two paradigmatic models with distinct natural bipartitions: the transverse-field Ising (TFI) chain (entanglement between modes $(k, -k)$ and the Su-Schrieffer-Heeger (SSH) model (entanglement between sublattices $(A,B)$ at the same $k$, as shown in Fig.~\ref{fig: entropy.otoc}, thereby establishing the universality of this information-theoretic characterization of DQPTs.

\textit{Transverse-field Ising model.---}
We begin by considering the paradigmatic one-dimensional TFI model \cite{Sachdev2011book, Suzuki2013book, Cao2024pra}, described by the Hamiltonian
\begin{equation}
    H = -J\sum_{n=1}^{N}\sigma_{n}^{z}\sigma_{n+1}^{z} - h\sum_{n=1}^{N}\sigma_{n}^{x},
\end{equation}
where $\sigma_{n}^{x,z}$ are the Pauli matrix, $J > 0$ denotes the ferromagnetic coupling, and $h$ represents the transverse field. Through the Jordan-Wigner transformation followed by Fourier transformation, the Hamiltonian decouples in momentum space into independent Bogoliubov-de Gennes (BdG) sectors. For each $k>0$, we introduce the Nambu spinor $\hat{\Psi}_{k} = (c_{k}, c_{-k}^{\dag})^{T}$, yielding
\begin{equation}
    H = \sum_{k>0}\hat{\Psi}_{k}^{\dag}\mathcal{H}_{k}\hat{\Psi}_{k}, 
\end{equation}
with the Bogoliubov-de Gennes matrix
\begin{equation}
    \mathcal{H}_k = \begin{pmatrix}
        -J\cos k - h & iJ\sin k \\
        -iJ\sin k & J\cos k + h
    \end{pmatrix}.
\end{equation}
Diagonalization is achieved via a Bogoliubov rotation $c_{k} = \cos{\theta_{k}}\eta_{k} + i\sin{\theta}_{k}\eta_{-k}^{\dag}$, with $\tan{2\theta_{k}} = J\sin{k}/(h+\cos{k})$. In terms of the quasiparticles $\eta_{k}$, the Hamiltonian takes the diagonal form $H = \sum_{k} \varepsilon_{k}( \eta_{k}^{\dagger}\eta_{k} - \tfrac{1}{2})$, where the dispersion reads $\varepsilon_{k} = \sqrt{(h+\cos k)^{2} + J^{2}\sin^{2}{k}}$.

We investigate nonequilibrium dynamics following a sudden quench: the system is initialized in the ground state $|\psi(0)\rangle$ of $H(h_{0})$, corresponding to the vacuum of the pre-quench quasiparticles $\eta_{k}(h_{0})$. At $t=0$, the transverse field is abruptly switched to $h_{1}$, and the state evolves unitarily under the post-quench Hamiltonian $\tilde{H}=H(h_{1})$. In each $(k, -k)$ subspace, the time-evolved state can be expressed in the post-quench Bogoliubov basis as 
\begin{equation}
    |\psi_{k}(t)\rangle = \alpha_{k}(t)|\tilde{0}_{k}\tilde{0}_{-k}\rangle + \beta_{k}(t) |\tilde{1}_{k}\tilde{1}_{-k}\rangle
\end{equation}
where the coefficients are given by $\alpha_{k}(t) = \cos{\Delta\theta_{k}}e^{i\tilde{\varepsilon}_{k}t}$ and $\beta_{k}(t) = \sin{\Delta\theta_{k}}e^{-i\tilde{\varepsilon}_{k}t}$, with $\Delta\theta_{k}=\tilde{\theta}_{k}-\theta_{k}$ measuring the difference between the pre- and post-quench Bogoliubov angles.

\textbf{A. Momentum-space entanglement entropy.}

To quantify the entanglement between the $k$ and $-k$ modes, we first write the full density matrix $\rho_{k}$ in the basis $\{|\tilde{0}_{k}\tilde{0}_{-k}\rangle, |\tilde{1}_{k}\tilde{0}_{-k}\rangle, |\tilde{0}_{k}\tilde{1}_{-k}\rangle, |\tilde{1}_{k}\tilde{1}_{-k}\rangle\}$, which reads
\begin{equation}
    \rho_{k,-k}(t) = \begin{pmatrix}
        |\alpha_{k}(t)|^{2} & 0  & 0 & \alpha_{k}(t)\beta_{k}^{*}(t) \\
        0 & 0 & 0 & 0 \\
        0 & 0 & 0 & 0 \\
        \alpha_{k}^{*}(t)\beta_{k}(t) & 0 & 0 & |\beta_{k}(t)|^{2}
    \end{pmatrix}.
\end{equation}
Hence the reduced density matrix for the $k-$mode is constructed by tracing out it partner $-k$:
\begin{equation}
\rho_k = \operatorname{Tr}_{-k} \rho_{k,-k} = \operatorname{diag}\big(|\alpha_k|^2, |\beta_k|^2\big).
\end{equation}
Remarkably, the occupations are time-independent, reflecting the integrable nature of the model. The associated entanglement entropy is then
\begin{equation}
    \mathcal{S}_{k} = -\operatorname{Tr}\bigl[\rho_{k}\ln{\rho_{k}}\bigr] = -p_{k}\ln{p_{k}} - (1-p_{k})\ln{(1-p_{k})},
\end{equation}
where $p_{k} \equiv \cos^{2}{\Delta\theta_{k}}$. For a two-level system, the maximum possible entropy is $\mathcal{S}_{k,\mathrm{max}} = \ln{2}$, which is attained precisely when $p_{k}=1/2$, i.e., when 
\begin{equation}\label{eq: Ising.S.max}
    \cos^{2}{\Delta\theta_{k}} = \frac{1}{2} \Leftrightarrow \cos{2\Delta\theta_{k}} = 0.
\end{equation}

\textbf{B. Loschmidt amplitude and DQPT condition}

The Loschmidt amplitude factorizes into momentum sectors \cite{Heyl2013prl, Karrasch2013prb}, $\mathcal{G}(t) = \prod_{k>0} \mathcal{G}_k(t)$ with $\mathcal{G}_k(t) = \langle\psi_k(0)|\psi_k(t)\rangle$. A straightforward calculation gives the Loschmidt echo per mode,
\begin{equation}
    \mathcal{L}_{k}(t)=|\alpha_{k}(t)|^{4}+|\beta_{k}(t)|^{4}+2|\alpha_{k}(t)|^{2}|\beta_{k}(t)|^{2}\cos{2\tilde{\varepsilon}_{k}t}.
\end{equation}
A DQPT occurs when there exists at least one critical momentum $k^{*}$ and a critical time $t^{*}$ such that $\mathcal{L}_{k^{*}}(t^{*})=0$. This condition requires simultaneously 
\begin{equation}
    |\alpha_{k}(t)|^{2}=|\beta_{k}(t)|^{2}=\frac{1}{2},~ \text{and}~\cos{2\tilde{\varepsilon}_{k}t}=-1.
\end{equation}
The first equality is exactly condition (\ref{eq: Ising.S.max}) for the entanglement entropy to reach its maximum. Thus, in the TFI chain, the onset of a DQPT is universally accompanied by the saturation of entanglement in the corresponding momentum mode.
in the initial state 

\textbf{C. Momentum-space OTOC.}

To probe information dynamics within the same entangled mode pair, we compute the momentum-resolved OTOC defined in Eq.~(\ref{eq: otoc.definition}) with the choice $V=n_{-k}$ and $W(t)=n_{k}(t)$in the initial state 
, where $n_{k}=c_{k}^{\dag}c_{k}$ is the occupation-number operator. After expressing the operators in the initial-state basis and performing the fermionic algebra, we obtain the closed-form result
\begin{equation}\label{eq: Ising.otoc}
    C_{k}(t) = -\langle[n_{k}(t), n_{-k}]^{2}\rangle = \sin^{2}{(2\tilde{\theta}_{k})}\sin^{2}{(\tilde{\varepsilon}_{k}t)}\cos^{2}{2\Delta\theta_{k}}.
\end{equation}

Several key features are immediate:
\begin{enumerate}
    \item $C_{k}(t)\geq0$,  as required for an OTOC of Hermitian operators.
    \item The time dependence is periodic with frequency with $2\tilde{\varepsilon}_{k}$, reflecting the free-particle dynamics.
    \item Most importantly, the OTOC vanishes identically for all times, $C_{k}(t)\equiv0$, precisely when $\cos{(2\Delta\theta_{k})}=0$---the same condition (\ref{eq: Ising.S.max}) that maximizes the entanglement entropy $\mathcal{S}_{k^{*}}$ and enables a Fisher zero in the Loschmidt amplitude.
\end{enumerate}

This establishes the exact equivalence between maximal mode entanglement and complete suppression of information scrambling within the $(k, -k)$ pair. For a critical momentum $k^{*}$ satisfying $\cos{(2\Delta\theta_{k^{*}})}=0$, not only does the entanglement entropy saturate to $\ln{2}$, but the OTOC remains zero throughout the evolution, indicating a persistent freezing of quantum information between the two modes. The DQPT critical point $(k^{*}, t^{*})$ is the special instance where this frozen, maximally entangled channel also satisfies the dynamical phase condition $\cos{2\tilde{\varepsilon}_{k}t}=-1$, leading to a nonanalyticity in the global Loschmidt rate.

\textit{SSH chain.---}
We next turn to the SSH model, a prototypical two-band topological insulator \cite{Su1979prl, Asboth2016book, Cao2025pra}. Its Hamiltonian reads
\begin{equation}
    H = \sum_{n=1}^{N}\bigl(t_{1}c_{A,n}^{\dag}c_{B,n} + t_{2}c_{B,n}^{\dag}c_{A,n+1} + h.c. \bigr),
\end{equation}
where $c_{A,n}, c_{B,n}$ annihilate a ferimion on sublattices $A$ and $B$ of unit cell $n$, $t_{1}$ and $t_{2}$ denote the intra-cell and inter-cell hoppings, respectively. After Fourier transformation, the Bloch Hamiltonian in the sublattice basis $\hat{\Psi}_{k}=(c_{A,k}, c_{B,k})^{T}$ takes the form
\begin{equation}
    \mathcal{H}_{k} = \begin{pmatrix}
        0 & t_{1}+t_{2}e^{-ik} \\
        t_{1}+t_{2}e^{ik} & 0
    \end{pmatrix} = \textbf{d}_{k} \cdot \vec{\sigma},
\end{equation}
with $\textbf{d}_{k} = (t_{1}+t_{2}\cos{k}, t_{2}\sin{k}, 0)$. The spectrum consists of two bands $E_{\pm}(k) = \pm|\textbf{d}_{k}|$. It is convenient to introduce a polar angle $\theta_{k}$ in the $xy-$plane via
\begin{equation}
    \tan{\theta_{k}} = \frac{t_{2}\sin{k}}{t_{1}+t_{2}\cos{k}},
\end{equation}
so that $\textbf{d}_{k} = |\textbf{d}_{k}|(\cos{\theta_{k}}, \sin{\theta_{k}}, 0)$. Diagonalization is achieved by the unitary rotation
\begin{equation}
    U_k = \begin{pmatrix}
        \cos{\frac{\theta_{k}}{2}} & -\sin{\frac{\theta_{k}}{2}} \\
        \sin{\frac{\theta_{k}}{2}} & \cos{\frac{\theta_{k}}{2}}
    \end{pmatrix}, U_{k}^{\dag}\mathcal{H}_{k}U_{k}=\mathrm{diag}[E_{+}(k), E_{-}(k)].
\end{equation}
The new operators $\eta_{\pm,k}=U_{k}^{\dag}\Psi_{k}$ describes the eigenmodes, and the Hamiltonian becomes
\begin{equation}\label{eq: SSH.diagonal}
    H = \sum_{k}[E_{+}(k)\eta_{+,k}^{\dag}\eta_{+,k} + E_{-}(k)\eta_{-,k}^{\dag}\eta_{-,k}].
\end{equation}
The ground state of the half-filled system corresponds to filling the lower band, which in the sublattice occupation basis $|n_{A,k}n_{B,k}\rangle, (n_{A,k}, n_{B,k}=0,1)$ reads
\begin{equation}
    |\psi_{-,k}\rangle = \sin{\frac{\theta_{k}}{2}}|10\rangle + \cos{\frac{\theta_{k}}{2}}|01\rangle.
\end{equation}

We now consider a quantum quench: the system is initialized in the ground state of $H(t_{1}, t_{2}^{i})$, at $t=0$, the intercell coupling is abruptly changed to $t_{2}^{f}$. The time‑evolved state in each momentum sector can be expanded in the post‑quench eigenbasis as
\begin{equation}
    |\psi_{k}(t)\rangle = p_{+}(k)e^{-iE_{+}(k)t}|\psi_{+,k}^{f}\rangle + p_{-,k}e^{-iE_{-}(k)t}|\psi_{-,k}^{f}\rangle,
\end{equation}
with overlap coefficients
\begin{align}
    p_{+,k} &= \langle\psi_{+,k}^{f}|\psi_{-,k}^{i}\rangle = -\sin\frac{\Delta\theta_{k}}{2}, \\
    p_{-,k} &= \langle\psi_{-,k}^{f}|\psi_{-,k}^{i}\rangle = -\cos\frac{\Delta\theta_{k}}{2},
\end{align}
where $\Delta\theta_{k}=\theta_{k}^{f}-\theta_{k}^{i}$. 

\textbf{A. Sublattice momentum-space entanglement entropy.}

In the occupation basis, the full density matrix for $(A, B)$ pair at momentum $k$ is
\begin{equation}
    \rho_{k} = \begin{pmatrix}
        0 & 0 & 0 & 0 \\
        0 & \cos^{2}{\frac{\Delta\theta_{k}}{2}} & \sin{\frac{\Delta\theta_{k}}{2}}\cos{\frac{\Delta\theta_{k}}{2}} & 0 \\
        0 & \sin{\frac{\Delta\theta_{k}}{2}}\cos{\frac{\Delta\theta_{k}}{2}} & \sin^{2}{\frac{\Delta\theta_{k}}{2}} & 0 \\
        0 & 0 & 0 & 0
    \end{pmatrix}.
\end{equation}
Crucially, unlike the Ising case where entanglement is defined between $(k,-k) $momentum partners, here the relevant bipartition is between the two sublattices $A$ and $B$ at the same momentum. Tracing out the $B$ sublattice yields the reduced density matrix for the $A$ sublattice:
\begin{equation}
    \rho_{A,k} = \operatorname{Tr}_{B} \rho_{k} = \operatorname{diag}\big(\cos^{2}{\frac{\Delta\theta_{k}}{2}}, \sin^{2}{\frac{\Delta\theta_{k}}{2}}\big).
\end{equation}
The corresponding sublattice momentum‑space entanglement entropy is therefore
\begin{equation}
    \mathcal{S}_{k} = -\operatorname{Tr}\bigl[\rho_{A,k}\ln{\rho_{A,k}}\bigr] = -q_{k}\ln{q_{k}} - (1-q_{k})\ln{(1-q_{k})},
\end{equation}
where $q_{k} \equiv \cos^{2}{\frac{\Delta\theta_{k}}{2}}$. This entropy is maximized, $\mathcal{S}_{k,\mathrm{max}} = \ln{2}$, when the two outcomes are equally probable:
\begin{equation}\label{eq: SSH.S.max}
    \cos^{2}{\frac{\Delta\theta_{k}}{2}} \Leftrightarrow \cos{\Delta\theta_{k}} = 0.
\end{equation}

\textbf{B. Loschmidt amplitude and DQPT condition.}

The Loschmidt amplitude factorizes as  $\mathcal{G}(t)=\prod_{k}\mathcal{G}_{k}(t)$ with \cite{vajna2015prb}
\begin{equation}
    \mathcal{G}_{k}(t) = \cos{[E_{+}^{f}(k)^{f}t]} + i\hat{\textbf{d}}_{k}^{i}\cdot\hat{\textbf{d}}_{k}^{f}\sin{[(E_{+}^{f}(k)^{f}t]},
\end{equation}
where $\hat{\textbf{d}}_{k}= \textbf{d}_{k}/|\textbf{d}_{k}|$. Fisher zeros, and hence DQPTs, occur when there exists a critical $k^{*}$ and time $t^{*}$ such that $\mathcal{G}_{k^{*}}(t^{*}) = 0$. This requires simultaneously
\begin{equation}
    \hat{\textbf{d}}_{k}^{i}\cdot\hat{\textbf{d}}_{k}^{f}=0, ~\text{and}~\cos{[E_{+}^{f}(k)^{f}t]}=-1.
\end{equation}
The orthogonality condition $\hat{\textbf{d}}_{k}^{i}\cdot\hat{\textbf{d}}_{k}^{f}=0$ is equivalent to $\cos{\Delta\theta_{k}}=0$, which which is precisely condition (\ref{eq: SSH.S.max}) for maximal sublattice entanglement.

\textbf{C. Sublattice momentum-space OTOC.}

To probe information scrambling between the two sublattices at the same momentum, we compute the OTOC with the natural choice $V=n_{Bk}$ and $W(t)=n_{Ak}(t)$, where $n_{Ak}=c_{Ak}^{\dag}c_{Ak}$ and $n_{Bk}=c_{Bk}^{\dag}c_{Bk}$. Following a derivation analogous to that for the TFI model, we obtain
\begin{equation}\label{eq: SSH.otoc}
    C_{k}(t) = -\langle[n_{Ak}(t), n_{Bk}]^{2}\rangle = \sin^{2}{\theta_{k}^{f}}\sin^{2}{(|\textbf{d}_{k}^{f}|t)}\cos^{2}{\Delta\theta_{k}}.
\end{equation}
The structure closely parallels the TFI result (\ref{eq: Ising.otoc}): a non-negative product of a time-oscillatory factor $\sin^{2}{(|\textbf{d}_{k}^{f}|t)}$, a structure factor $\sin^{2}{\theta_{k}^{f}}$ encoding the post-quench band geometry, and the key quench-induced factor $\cos^{2}{\Delta\theta_{k}}$. Remarkably, when $\cos{\Delta\theta_{k}} = 0$---the same condition that maximizes $\mathcal{S}_{k}$ and enables a Fisher zero—the OTOC vanishes identically for all times, $C_{k}(t)\equiv0$. This establishes the exact correspondence between maximal sublattice entanglement and complete suppression of information scrambling between sublattices at the critical momentum $k^{*}$.

\textit{Conclusion.---}
In summary, we have established a unified, mode-resolved information-theoretic framework that fundamentally recasts the understanding of dynamical quantum phase transitions in translation-invariant free-particle systems. By introducing and analytically investigating the momentum-space entanglement entropy $\mathcal{S}_{k}(t)$ and a momentum-resolved OTOC $C_{k}(t)$, we have demonstrated that DQPTs are universally characterized by a triad of equivalent critical signatures at a critical momentum $k^{*}$: the emergence of a Fisher zero in the Loschmidt amplitude, the saturation of bipartite entanglement to its maximum value $\ln{2}$, and the complete, persistent suppression of information scrambling ($C_{k^{*}}(t) \equiv 0$).

This equivalence, rigorously proved in both the TFI and SSH models—despite their distinct natural bipartitions—reveals a profound and universal principle. At a DQPT, the critically entangled sector reaches a dynamical fixed point in information space: quantum correlations are maximized, and information flow between the entangled degrees of freedom is frozen. Our work thus elevates entanglement entropy and the OTOC from mere diagnostics to central dynamical order parameters for nonequilibrium quantum criticality. They provide a direct, mode-resolved window into the quantum-informational restructuring that underlies the global nonanalyticity captured by the Loschmidt echo.

Looking forward, our framework opens several compelling avenues. The intimate link between maximal entanglement and arrested scrambling invites investigation in more complex settings, such as systems with long-range hoppings, higher dimensions, non-Hermitian dynamics \cite{Chen2022prb}, or beyond-integrable interactions, where mode coupling could qualitatively alter this relationship. Furthermore, it poses a concrete experimental challenge: to measure these momentum-space information-theoretic quantities in quantum simulators—such as cold atoms in optical lattices or superconducting qubit arrays—which already offer unprecedented control over momentum-resolved correlations \cite{Carcy2019prx, Florian2014science, Yu2025NatPhys}.

Ultimately, this work elevates entanglement and information scrambling from supporting observables to central dynamical order parameters in the story of nonequilibrium quantum phases. By tying DQPTs directly to the saturation of quantum correlations and the freezing of information flow, we provide a unified and intuitive understanding of their origin—one that is rooted in the fundamental information-theoretic structure of quantum many-body dynamics.

\begin{acknowledgments}
  K.C. was funded by Basic Research Program of Jiangsu (Grant No.~BK20250886).  J.W. was supported by the National Natural Science Foundation of China (Grant No.~11875047). S.C. was supported by National Key Research and Development Program of China (Grant No.~2021YFA1402104) and the National Natural Science Foundation under Grants No.~12474287 and No.~T2121001.
\end{acknowledgments}

\bibliography{reference}

\end{document}